\documentstyle{article}
\begin{document}
\textwidth=140mm
\textheight=200mm
\begin{center}
{\Large\bf
Formation time effect on $J/\psi$ dynamical nuclear suppression} \\
\bigskip
Sa Ben-Hao$^{1,2,4}$, Tai An$^{1,3}$, Wang Hui$^2$, and Liu Feng-He$^2$ \\
\begin{tabbing}
ttttt \= tt \= \kill
\>1. CCAST (World Lab.), P. O. Box 8730 Beijing, China. \\
\>2. China Institute of Atomic Energy, P. O. Box 275 (18), \\
\> \>Beijing, 102413 China.\footnotemark \\
\>3. Institute of High Energy Physics, Academic Sinica, \\
\> \> P. O. Box 918, Beijing, 100039 China.\\
\>4. Institute of Theoretical Physics, Academic Sinica, \\
\> \>Beijing China.\\
\end{tabbing}
\end{center}
\footnotetext{mailling address. \\
Email: sabh@mipsa.ciae.ac.cn; taian@hptc1.ihep.ac.cn}

\normalsize
\begin{abstract}
The proposed hadronic and string cascade model, JPCIAE, for 
ultrarelativistic nucleus - nucleus collisions based on the LUND model and the
PYTHIA event generator especially, is used to investigate the $J/\psi$
suppression due to the nuclear absorption of a $J/\psi$ in minimum bias pA and 
BA collisions at 200 A GeV energy.
With the different sets of reasonable formation time for hadron and $J/\psi$
the results of $J/\psi$ suppression factor from both the usual scenario
and the Glauber - like simulations are comparable with all the NA38 pA and BA 
data, except the NA50 data of Pb + Pb collisions. However,
the difference between the usual scenario and the Glauber - like
simulation, hence the difference between the dynamical simulation and Glauber
theory, can not be ignored. The sensitive effect of the hadron formation time
on the $J/\psi$ suppression is studied in detail. The results seem to denote
that for the $J/\psi$ suppression the meson absorption plays role in pA as 
well as in BA collisions. 

\end{abstract}
\baselineskip 0.8cm
\parindent=0.3cm
\parskip=0.3cm
\hspace{0.3cm}

More than ten years ago Matsui and Satz [1] suggested that the
suppression of $J/\psi$ yield in relativistic nucleus - nucleus collisions
might be a powerful signature for the QGP formation. Since then, a number of 
corresponding experiments have been stimulated [2 - 4] to measure the $J/\psi$ 
yield via its dimuon decay. A significant suppression of the $J/\psi$ yield 
from pA collisions to BA collisions has already been observed in these 
experiments. So far, except the anomalous nuclear suppression observed in Pb + 
Pb reactions at 158 A GeV/c [4] the normal suppression has been well explained 
within the Glauber theory (absorption model) [5 - 14]. However, the mechanism 
for the anomalous suppression in Pb + Pb collisions is still a debating issue 
[9 - 15].

Recently in [16 - 17] a covariant transport approach has been used to
investigate both the normal and anomalous nuclear suppression of the $J/\psi$
yield. They concluded that the data of $J/\psi$ suppression from pA to BA
collisions, including Pb + Pb at 158 A GeV/c, can be described without assuming
the formation of QGP in these collisions. However, not only their conclusions
need to have more response using dynamical models but also the dynamical
ingredients, such as the formation time effect, the role of meson, etc.,
need to have more study.

In this letter we propose a hadronic and string cascade model, JPCIAE, for 
relativistic nucleus - nucleus collisions based on the LUND model and the
PYTHIA event generator [18]. We first inspect this model and the corresponding
event generator via comparing with the NA35 data of the negative charge
multiplicity, the rapidity and transverse momentum distributions of the
negative charge particles and the participant protons in  pp, pA, and BA
collisions [19 - 20]. The model and the corresponding event
generator are then used to investigate the $J/\psi$ nuclear suppression and
the effects of dynamical ingredients. The results seem to declare that all the
NA38 data of $J/\psi$ normal suppression from pA to BA collisions [2 - 3] can
be fairly described by this model, except the NA50 data of Pb + Pb 
collisions [4]. It is also shown that the formation time of produced particles 
has very sensitive effect on the $J/\psi$ suppression and it turns out that 
the meson absorption plays a similar role in pA and BA collisions. We point 
out also the difference between two scenarios $-$ the complete rescattering of 
produced hadrons (named as usual scenario) and the rescattering only between a 
$J/\psi$ and other produced hadrons (named as Glauber-like).

In JPCIAE the simulation is performed in the laboratory system. The origin 
of coordinate space is positioned at the center of the target nucleus
and the beam direction is taken as the z axis. As for the origin of time
it is set at the moment when the distance between the projectile and target 
nucleus along z direction is equal to zero (collision time can be negative).

The colliding nucleus is depicted as a sphere with radius $\sim$ 1.05 A$^{1/3}$
($A$ refers to the mass number of a nucleus) in its rest frame. The
spatial distribution of nucleons in this frame is sampled randomly 
due to the Woods - Saxon distribution. The projectile
nucleons are assumed to have an incident momentum per nucleon and the target
nucleons are at rest. That means the Fermi motion in a nucleus and the mean
field of a nuclear system are here neglected due to interest of relativistic
energy in question. For the distribution of the projectile nucleons the Lorentz 
contraction is taken into account.

Then the collision time is calculated according to the requirement 
that the minimum approaching distance of a colliding pair should be 
less or equal to the value $\sqrt{\sigma_{tot}/\pi}$, where $\sigma_{tot}$ is 
the total cross section of the colliding pair. The initial collision time list 
is composed of the colliding nucleon pairs, in each pair here one partner is 
from the projectile nucleus and the other from the target nucleus.

If the CMS energy of a colliding pair (hadron - hadron collision), selected 
due to the least collision time from the collision time list, is larger than 
or equal to $\sim$ 4 GeV two string states are formed and PYTHIA is called to 
provide the produced hadrons (scattered state), no string state is formed and 
the conventional scattering process [21 - 23] is executed otherwise.
Both of the particle list and the collision time list are then updated and
they are now not only composed of the projectile and target nucleons but also
the produced hadrons. The history of an event ends when the collision time list
is empty.

In PYTHIA a lot of parton - parton (parton refers to quark or gluon) QCD
processes have been considered, including the $J/\psi$ production,
\begin{equation}
g + g \rightarrow J/\psi + g (1.3\times 10^{-5} mb).
\end{equation}
A user is allowed to run the program with any desired subset of these processes.
We have devised a switch to turn over from the preprogrammed channel to the
$J/\psi$ production channel. It is worthy pointing out that any operation with
desired subset of the processes by user, including the $J/\psi$ production
channel defined here, is a kind of bias sampling, which enhances the
probabilities of those desired processes. In order to overcome the
corresponding bias the selection of calling PYTHIA or of executing conventional
scattering for each hadron - hadron collision is decided further according to
the probability which is equal to the parametrized $J/\psi$ production cross 
section [24]
\begin{equation}
\sigma_{NN\rightarrow J/\psi +X} = d(1-\frac{c}{\sqrt {s}})^{12}
\end{equation}
(with c=3.097 GeV, d=2.37/B$_{\mu \mu}$ nb, B$_{\mu \mu}$ = 0.0597 is the
branching ratio of the $J/\psi$ to dimuons) multiplied by a factor. That factor
is adjusted so that the number of $J/\psi$ produced in each simulating event is
around one, the same as in the experiment [2].

One more point needed to be mentioned here is that in the original JETSET 
program, which deals with the fragmentation of a string and runs together with 
PYTHIA, the leading particle in a nucleon - nucleon collision is assumed to 
carry about half of the incident energy. But the experiments of nucleus - 
nucleus collisions at relativistic energies reveal that a incoming nucleon 
loses a smaller fraction of its energy in each binary nucleon - nucleon 
collision except its last collision with the target nucleon, in which it loses 
about half of its energy, and a stopping law is proposed in [25 - 26] to 
handle this situation. We in this program have also applied the stopping law 
to calculate the energy fraction that a leading particle takes after each 
binary nucleon - nucleon collision.

For inspecting the model and the corresponding program (event generator) we
first compare the calculated (using preprogrammed channel) negative charge
multiplicity, the rapidity and transverse momentum distributions for the
negative charge particles and for the participant protons in  pp, pA, and BA
collisions with the corresponding data [19 - 20]. The comparisons of negative 
charge multiplicity for pp and pA reactions are shown in Tab. 1 and for BA in 
Tab.2 . Fig. 1 gives the comparison for the rapidity distributions of the 
negative charge particles in central S + S and N + N minimum bias collisions 
(upper frame) and for the participant protons in S + S central and peripheral 
collisions (lower frame). The transverse momentum distributions in central S + 
S collisions for the negative charge particles and for the participant protons 
are given in the upper frame and the lower frame of Fig. 2, respectively. One 
sees from these tables and figures that the agreement between theory and 
experiment is reasonably good.

We are then turning to the calculations with $J/\psi$ production channel. 
Since the purpose of this letter is to explore the physics behind the NA38 and 
NA50 data and not to fit the data as good as possible. We first fix two 
reasonable sets of parameters to calculate the $J/\psi$ suppression factors in 
minimum bias pA and BA collisions at the scaled energy 200 A GeV and compare 
them with the corresponding data in Fig. 3. The experimental $J/\psi$ 
suppression factor is defined as [17]
\begin{equation}
S_{exp.}^{J/\psi} = (\frac{B_{\mu\mu}\sigma_{BA}^{J/\psi}}{\sigma_{BA}^{DY,2.9-
  4.5 GeV}})/(\frac{B_{\mu\mu}\sigma_{pd}^{J/\psi}}{\sigma_{pd}^{DY}}),
\end{equation}
since the $J/\psi$ yield is measured via its dimuon decay and the Drell - Yan
provides the background of the dimuon invariant mass spectrum. As for the
theoretical definition of the $J/\psi$ suppression factor it is expressed as 
[17, 27]
\begin{equation}
S_{theo.}^{J/\psi} = \frac{M_{J/\psi}}{M_{J/\psi}(0)}
\end{equation}
where $M_{J/\psi}(0)$ refers to the multiplicity of primary $J/\psi$ and
$M_{J/\psi}$ to the multiplicity of $J/\psi$ after final interactions. The 
open circles with error bar in Fig. 3 are the experimental data (cited 
directly from [27]). In Fig. 3 the full circles are the results of usual 
scenario calculations with parameter set 1: the meson formation time $\tau_M$ 
= 1.2 fm/c and the $J/\psi$ formation time $\tau_{J/\psi}$ = $f_{J/\psi}\times 
\tau_M$, $f_{J/\psi}$ = 0.5, the full triangles are the results of Glauber - 
like calculations with parameter set 2: $\tau_M$ = 0.8 fm/c, and $f_{J/\psi}$ 
= 0.5, and the open triangles are the results of Glauber - like calculations 
with parameter set 1. As for the nucleon formation time it is assumed to be 
zero. Since the cross section of $J/\psi$ - hadron interaction is still an 
open problem [28 - 29], we do not address this question here and adopt simply 
the values: $\sigma_{J/\psi-B}^{Abs}$ = 6 mb and $\sigma_{J/\psi-M}^{Abs}$ = 3 
mb as usual. The corresponding total cross section used in the program are 
$\sigma_{J/\psi-B}^{tot}$ = 7.2 mb and $\sigma_{J/\psi-M}^{tot}$ = 4.0 mb. The 
following reactions of $J/\psi$ with B (baryon) and M (meson) are considered
\begin{eqnarray}
J/\psi + B \rightarrow \Lambda_c + \bar{D} ,\\
J/\psi + M \rightarrow D + \bar{D} .
\end{eqnarray}

One sees from Fig. 3 that both the results of the usual scenario with parameter
set 1 and the results of Glauber - like calculations with parameter set 2, 
i.e. the full circles and the full triangles, are comparable with all the
experimental data, except the NA50 data of Pb + Pb reactions. However, the
Glauber - like calculation needs to have a smaller meson formation time than
the usual scenario, as the freeze - out time of a hadron in the usual scenario 
is longer than ones in the Glauber - like situation. That is because the
reinteraction between hadrons (except $J/\psi$) is not taken into account in
the Glauber - like situation. Comparing the results of usual scenario with 
the results of Glauber - like calculation using the same parameter set 1 
(cf. the full circles and the open triangles in Fig. 3) one knows that the 
difference between the usual scenario and Glauber - like situation and then 
between the dynamical simulation and the Glauber theory should not be ignored. 
This conclusion is in consistent with [16 - 17]. Of course, the 
difference between the usual scenario and Glauber - like situation is also 
formation time dependent.

Tab. 3 gives the $J/\psi$ suppression factor calculated for minimum bias p + 
Al, p + Cu, p + Ag, P + U, O + Cu, and O + U collisions at 200 A GeV under the
usual scenario and the parameters of $f_{J/\psi}$ = 0.5 and of various $\tau
_M$. Tab. 4 gives the results calculated for the same collisions
as the ones in Tab. 3 but with parameters of $\tau_M$ = 1.2 fm/c and of various
$f_{J/\psi}$. These tables indicate that the $J/\psi$ suppression factor is
very sensitive to the formation time of meson and $J/\psi$. At the same
formation time of meson and $J/\psi$ the $J/\psi$ suppression factor, both in 
pA and BA collisions, decreases with the increasing of target mass. In the 
case of 1 + $A^{1/3}$ (pA collision) $\simeq B^{1/3}$ + $A^{1/3}$ (BA 
collision, $B$ and $A$ here refer also 
to the mass numbers of projectile and target nuclei) the $J/\psi$ suppression 
factor in pA collision is larger than ones in BA collision, that is because of 
the less hadrons are produced in pA than in BA collisions.   

In order to check whether mesons play different role for the $J/\psi$
suppression in pA and BA collisions at the same incident energy and centrality,
a factor defined by
\begin{equation}
f_M = \frac{S_{theo.}^{without M} - S_{theo.}^{with M}}{S_{theo.}^{without M}}
\end{equation}
is introduced. The results of $f_M$ calculated for the same collisions as the 
ones in Tab. 3 under the usual scenario and the parameter set 1, are given in 
Tab. 5 . One sees from this table that in pA collisions the role of mesons 
in the $J/\psi$ suppression is not negligible and it is increased with the 
increasing of target mass. That is inconsistent with the conclusion in [6 - 13
, 16 - 17]. At the fixed mass of the projectile nucleus the effect of meson on 
the $J/\psi$ suppression depends mainly on the mass of the target nucleus both 
in pA and BA collisions.

In summary We have proposed a hadronic and string cascade model, JPCIAE, for 
ultrarelativistic nucleus - nucleus collisions based on the LUND model and
the PYTHIA event generator especially. It has been used to investigate the
$J/\psi$ suppression in minimum bias pA and BA collisions at the scaled energy 
of 200 A GeV. With the different sets of reasonable formation time for hadron 
and $J/\psi$ the results of $J/\psi$ suppression factor from both the usual 
scenario and the Glauber - like simulations are comparable with all the NA38 pA 
and BA data, except the NA50 data of Pb + Pb collisions. However, the 
difference between the usual scenario and the Glauber - like simulations, 
hence the difference between the dynamical simulation and Glauber theory, can 
not be ignored. The sensitive effect of the hadron formation time is studied 
in detail. Meanwhile, the results seem to denote that for the $J/\psi$ 
suppression mesons play role in pA as well as in BA collisions. 

The authors like to thank C. Y. Wong, Nu Xu, Guo - Qiang Li, and Bao - An Li
for valuable discussions. This work is supported both by the national Natural
Science Foundation of China and the Nuclear Industry Foundation of China.

\begin{center}Figure Captions\end{center}
\begin{quotation}
Fig. 1 The rapidity distributions for (a) h$^-$ and (b) participant proton.
The N+N data and corresponding results of JPCIAE have been multiplied by 10 for
the convenience of comparison. The labels are the experimental 
data and the curves are the corresponding results of JPCIAE.

Fig. 2 The transverse momentum distributions for (a) h$^-$ and (b) participant
proton. The labels are the experimental data and the the curves are the
corresponding results of JPCIAE.

Fig. 3 The $J/\psi$ suppression factor versus the product of mass numbers of
the projectile and the target nuclei in  minimum bias pA and BA collisions at 
200 A GeV. See text for the detail.
\end{quotation}

\newpage
\begin{tabular}{cccc}
\multicolumn{4}{c}{Table 1. Negative charge multiplicity in pp}\\
\multicolumn{4}{c}{and mini bias pA collisions at 200 GeV/c}\\
\hline
\hline
       & p + p & p + S & p + Ag \\
\hline
NA35 data& 2.85$\pm$0.3& 5.7$\pm$0.2& 6.2$\pm$0.2 \\
JPCIAE   & 2.84        & 4.91       & 5.81         \\
\hline
\hline
\end{tabular}

\begin{tabular}{ccc}
\multicolumn{3}{c}{Table 2. Negative charge multiplicity}\\
\multicolumn{3}{c}{in central BA collisions at 200A GeV}\\
\hline
\hline
       & S + S & S + Ag \\
\hline
NA35 data& 98$\pm$3& 170$\pm$8  \\
JPCIAE   & 107        & 173    \\
\hline
\hline
\end{tabular}

\begin{tabular}{ccccccc}
\multicolumn{7}{c}{Table 3. $J/\psi$ suppression factor in the reactions of
                   p+Al, p+Cu, }\\
\multicolumn{7}{c}{p+Ag, p+U, O+Cu, and O+U at 200A GeV, $f_{J/\psi}$=0.5 }\\
\hline
\hline
 $\tau_M (fm/c) $ & p + Al & p + Cu & p + Ag & p + U & O + Cu & O + U  \\
\hline
 0.2      & 0.220  & 0.194  & 0.180  & 0.122 & 0.144  & 0.103  \\
 0.5      & 0.435  & 0.403  & 0.369  & 0.314 & 0.313  & 0.246  \\
 1.2      & 0.888  & 0.836  & 0.779  & 0.738 & 0.713  & 0.635  \\
\hline
\hline
\end{tabular}

\begin{tabular}{ccccccc}
\multicolumn{7}{c}{Table 4. $J/\psi$ suppression factor in the reactions of}\\
\multicolumn{7}{c}{p+Al, p+Cu, p+Ag, p+U, O+Cu, and O+U at 200A GeV, }\\
\multicolumn{7}{c}{$\tau_M$=1.2 fm/c}\\
\hline
\hline
 $f_{J/\psi}$ & p + Al & p + Cu & p + Ag & p + U & O + Cu & O + U  \\
\hline
 0.25         & 0.538  & 0.496  & 0.472  & 0.453 & 0.389  & 0.343  \\
 0.50         & 0.888  & 0.836  & 0.799  & 0.738 & 0.713  & 0.635  \\
 0.75         & 0.943  & 0.932  & 0.921  & 0.859 & 0.832  & 0.773  \\
\hline
\hline
\end{tabular}

\begin{tabular}{ccccccc}
\multicolumn{7}{c}{Table 5. The factor $f_M$ in the reactions of }\\
\multicolumn{7}{c}{p+Al, p+Cu, p+Ag, p+U, O+Cu, and O+U at 200A GeV, }\\
\multicolumn{7}{c}{$\tau_M$=1.2 fm/c and $f_{J/\psi}$=0.5 fm/c}\\
\hline
\hline
 reaction               & p + Al & p + Cu & p + Ag & p + U & O + Cu & O + U  \\
\hline
 $S_{theo.}^{without M}$ & 0.903  & 0.865  & 0.875  & 0.810 & 0.772  & 0.706  \\
 $S_{theo.}^{with M}$    & 0.888  & 0.836  & 0.799  & 0.738 & 0.713  & 0.635  \\
 $f_M$                  & 0.0166 & 0.0335 & 0.0869 & 0.0889& 0.0764 & 0.101  \\
\hline
\hline
\end{tabular}

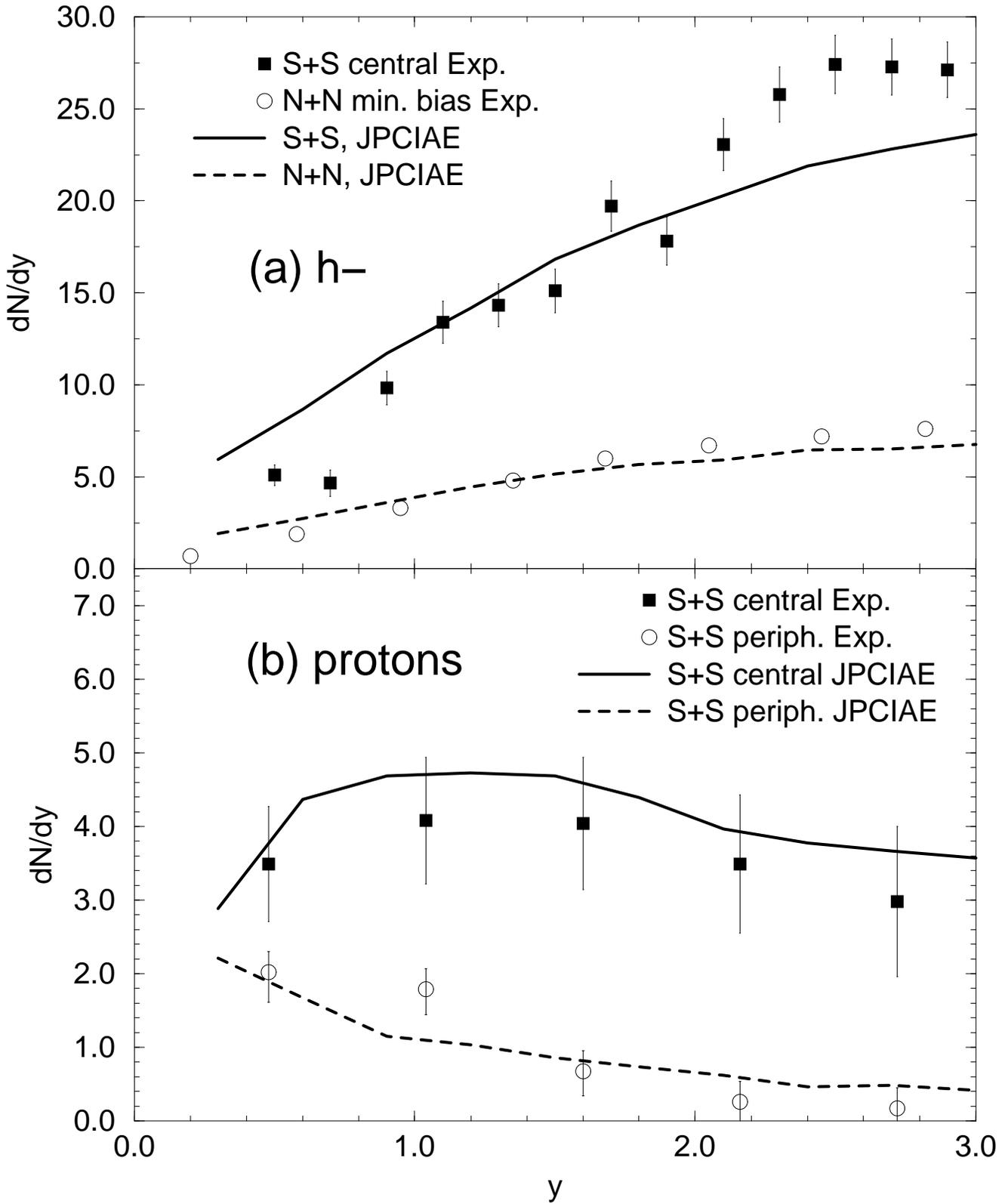
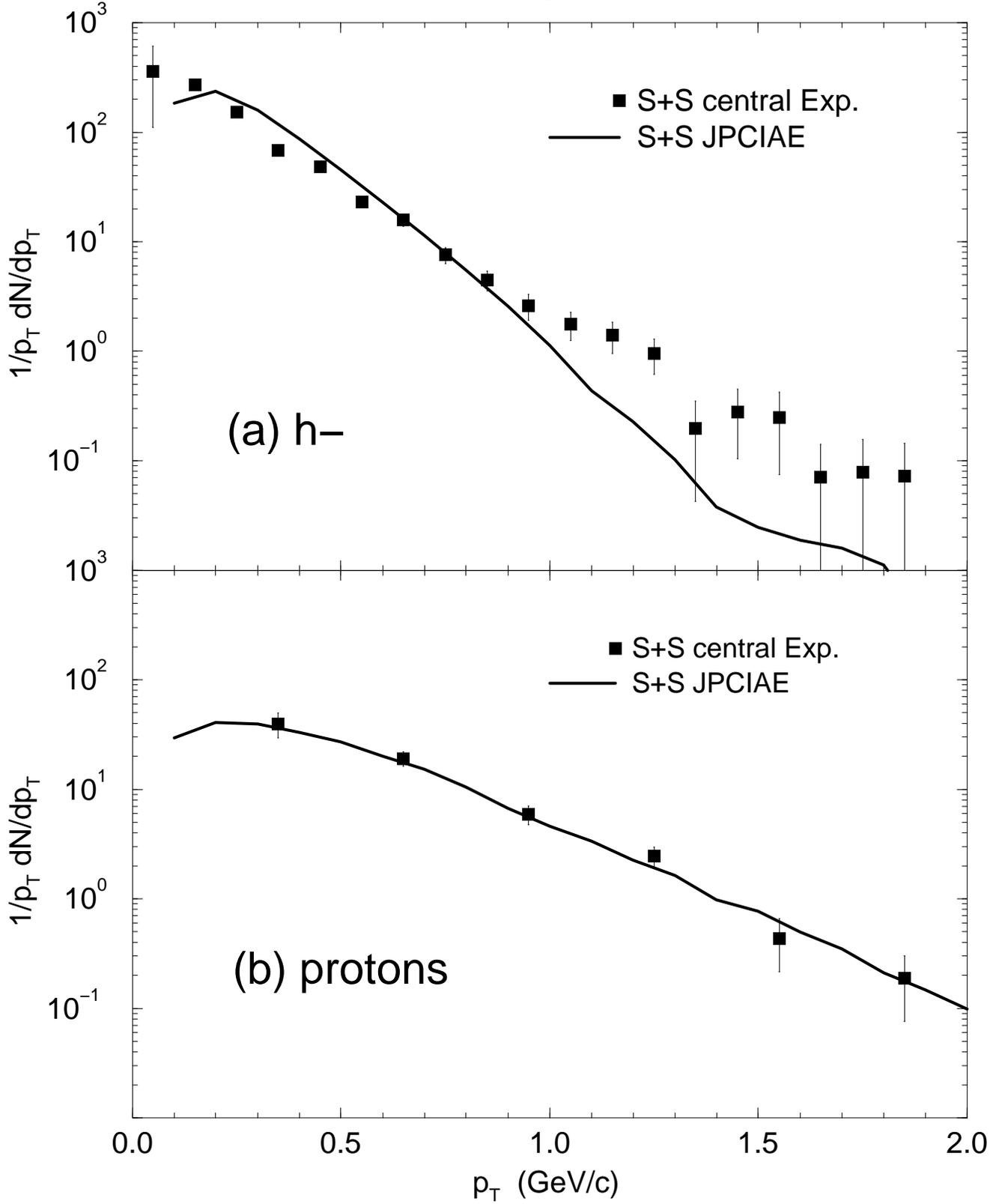
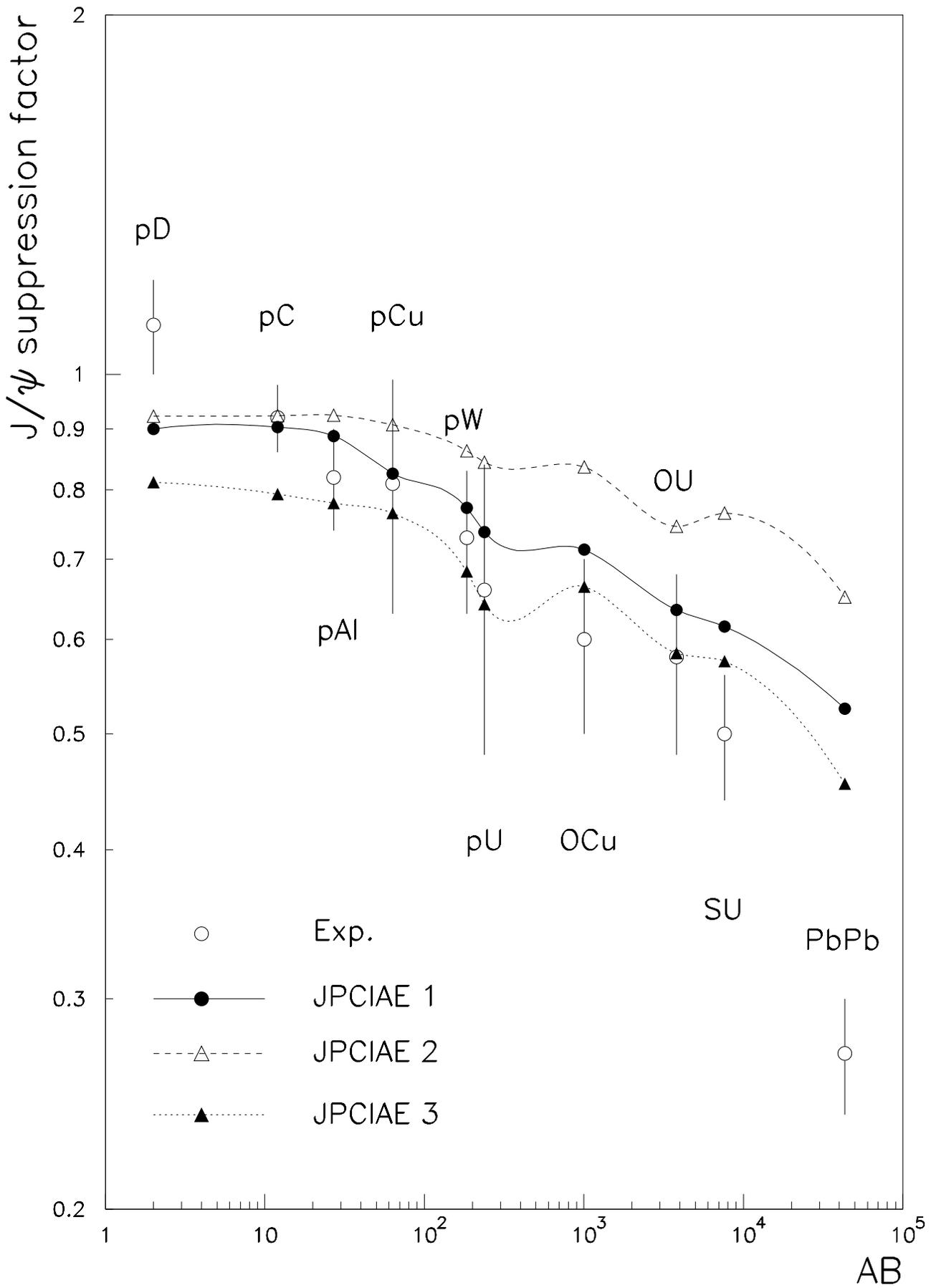
\end{document}